\documentclass[11pt]{article}

\usepackage{fullpage}

\usepackage{epsfig}
\usepackage{amssymb}
\usepackage{amsmath}
\usepackage{amsfonts}

\usepackage{pifont}

\begin{document}

\sloppy
\newcommand{\OMIT}[1]{} %
\newcommand{\jfootnote}[1]{} %
\newcommand{\efootnote}[1]{} %
\newcommand{\pfootnote}[1]{} %

\newcommand{\copelandalphaonescore}{\mathit{score}^1}
\newcommand{\copelandalphaone}{\mbox{\rm{}Copeland$^{1}$}}
\newcommand{\calS}{{\cal S}}
\newcommand{\pad}{{\rm{}Pad}}

\newcommand{\copelandalpha}{\mbox{\rm{}Copeland$^{\alpha}$}}
\newcommand{\copelandalphairrational}{\mbox{\rm{}Copeland$^{\alpha}_{\rm{}Irrational}$}}
\newcommand{\copelandalphascore}{\mathit{score}^\alpha}

\newenvironment{proofs}{\noindent{\bf Proof.}\hspace*{1em}}{\literalqed\bigskip}
\def\literalqed{{\ \nolinebreak\hfill\mbox{\qedblob\quad}}}

\newcommand{\startofproof}{\noindent{\bf Proof.}\hspace*{1em}}
\newcommand{\sproof}{\noindent{\bf Proof.}\hspace*{1em}}
\newcommand{\sproofof}[1]{\noindent{\bf Proof of {#1}.}\hspace*{1em}}
\newcommand{\eproofof}[1]{\noindent{\hspace*{0.1in} \hfil \hfill \mbox{\literalqed{} {#1}}}\quad\bigskip}

\newcommand{\condition}{\,\mid \:}

\newcommand{\sbribery}[1]{{\probbf \mbox{\rm{}#1}\hbox{-}\allowbreak\mbox{bribery}}}
\newcommand{\sdestbribery}[1]{{\probbf \mbox{\rm{}#1}\hbox{-}\allowbreak\mbox{destructive}\hbox{-}\allowbreak\mbox{bribery}}}

\newcommand\qedblob{\mbox{\ding{113}}}
\def\qedsymbol{{\ \nolinebreak\hfill\mbox{\qedblob\quad}}\smallskip}

\newcommand{\probbf}{\rm}
\newcommand{\wbribery}{{\probbf bribery}}
\newcommand{\wdbribery}{{\probbf {\dollars}bribery}}
\newcommand{\bribery}[2]{{\probbf \mbox{\rm{}#1}\hbox{-}\allowbreak\mbox{\rm{}#2}\hbox{-}\allowbreak{}bribery}}
\newcommand{\dbribery}[2]{{\probbf \mbox{\rm{}#1}\hbox{-}\allowbreak\mbox{\rm{}#2}\hbox{-}\allowbreak{\dollars}bribery}}
\newcommand{\sdbribery}[1]{{\probbf \mbox{\rm{}#1}\hbox{-}{\dollars}bribery}}
\newcommand{\udbribery}[2]{{\probbf \mbox{\rm{}#1}\hbox{-}\allowbreak\mbox{\rm{}#2}\hbox{-}\allowbreak{\dollars}bribery$_{\mbox{\rm{}unary}}$}}
\newcommand{\pudbribery}[2]{{\probbf \mbox{\rm{}#1}\hbox{-}\allowbreak\mbox{\rm{}#2}\hbox{-}\allowbreak{\dollars}bribery$'_{\mbox{\rm{}unary}}$}}
\newcommand{\sibribery}[1]{{\probbf \mbox{\rm{}#1\mbox{\Large$\rm{}_{Irrational}$}}\hbox{-}bribery}}
\newcommand{\sidestbribery}[1]{{\probbf \mbox{\rm{}#1\mbox{\Large$\rm{}_{Irrational}$}}\hbox{-}destructive\hbox{-}bribery}}
\newcommand{\spmicrobribery}[1]{{\probbf \mbox{\rm{}#1}\hbox{-}microbribery}}
\newcommand{\spdestmicrobribery}[1]{{\probbf \mbox{\rm{}#1}\hbox{-}destructive\hbox{-}microbribery}}
\newcommand{\wmanipulation}{\mbox{\probbf manipulation}}
\newcommand{\smanipulation}[1]{{\probbf \mbox{\rm{}#1}\hbox{-}\allowbreak{}manipulation}}
\newcommand{\manipulation}[2]{{\probbf \mbox{\rm{}#1}\hbox{-}\allowbreak\mbox{\rm{}#2}\hbox{-}\allowbreak{}manipulation}}
\newcommand{\scontrol}[2]{{\probbf \mbox{#1}\hbox{-}\mbox{\rm{}#2}}}
\newcommand{\sicontrol}[2]{{\probbf \mbox{#1}\mbox{\Large$\rm{}_{Irrational}$}\hbox{-}\mbox{\rm{}#2}}}
\newcommand{\sbcontrol}[2]{{\probbf \mbox{#1}\hbox{-}\ensuremath{k}\hbox{-}\mbox{\rm{}#2}}}

\newcommand{\naturals}{\mathbb{N}}
\newcommand{\integers}{\mathbb{Z}}

\newcommand{\cost}{\mathit{cost}}
\newcommand{\wincost}{\mathit{wincost}}
\newcommand{\tiecost}{\mathit{tiecost}}
\newcommand{\flowcost}{\mathit{flowcost}}
\newcommand{\flowvalue}{\mathit{flowvalue}}

\newcommand{\promote}{\mathit{promote}}
\newcommand{\demote}{\mathit{demote}}
\newcommand{\difference}{\mathit{diff}}
\newcommand{\diffAux}{\mathit{diffAux}}
\newcommand{\score}{\mathit{score}}
\newcommand{\llullscore}{\mathit{score}^\mathrm{L}}
\newcommand{\copelandscore}{\mathit{score}^\mathrm{C}}
\newcommand{\versus}{\mathrm{vs}}
\newcommand{\indegree}{\mathrm{deg}_{\mathrm{in}}}
\newcommand{\outdegree}{\mathrm{deg}_{\mathrm{out}}}

\newcommand{\electionrule}[1]{\emph{#1}}
\newcommand{\dollars}{{\probbf \$}}

\newcommand{\electionsystem}{\ensuremath{\cal E}}
\newcommand{\sssalpha}{\ensuremath{\alpha}}

\newcommand{\todo}[1]{\footnote{\textbf{SELFNOTE:} #1}}
\newcommand{\warn}[1]{\footnote{\textbf{MAY SHIFT IF BACK-SHIFT SUCCINCTNESS.} #1}}

\newcommand{\fpt}{\ensuremath{\mathrm{FPT}}}
\newcommand{\bigo}{{\protect\cal O}}
\newcommand{\bcj}{\mbox{\rm{}BC$_{j}$}}
\newcommand{\bvj}{\mbox{\rm{}BV$_{j}$}}
\newcommand{\card}[1]{{ \mathopen\parallel {#1} \mathclose\parallel }}

\newtheorem{theorem}{Theorem}[section]
\newtheorem{conjecture}[theorem]{Conjecture}
\newtheorem{corollary}[theorem]{Corollary}
\newtheorem{definition}[theorem]{Definition}
\newtheorem{lemma}[theorem]{Lemma}
\newtheorem{observation}[theorem]{Observation}
\newtheorem{fact}[theorem]{Fact}
\newtheorem{proposition}[theorem]{Proposition}
\newtheorem{example}[theorem]{Example}
\newtheorem{notation}[theorem]{Notation}
\newtheorem{trick}[theorem]{Trick Result}
\newtheorem{claim}[theorem]{Claim}
\newtheorem{construction}[theorem]{Construction}

\newcommand{\manyonereducesto}{\ensuremath{\leq_{\mathrm m}^{\mathrm p}}}
\newcommand{\dttreducesto}{\ensuremath{\leq_{\mathrm{dtt}}^{\mathrm p}}}

\newcommand{\p}{\ensuremath{\mathrm{P}}}
\newcommand{\np}{\ensuremath{\mathrm{NP}}}

\title{Copeland Voting Fully Resists
Constructive Control\thanks{%
Supported in part by DFG grant RO-1202/9-3 and RO-1202/11-1, NSF grants CCR-0311021,
CCF-0426761, and IIS-0713061, the Alexander von Humboldt Foundation's TransCoop
program, and a Friedrich Wilhelm Bessel Research Award. A version of this paper
also appears as URCS-TR-2007-923.
}
}

\author{
{Piotr Faliszewski} \\
Department of Computer Science\\ University of Rochester\\
Rochester, NY 14627
\and
{Edith Hemaspaandra}\thanks{Work done in part while visiting
Heinrich-Heine-Universit\"at D\"usseldorf.}\\
Department of Computer Science\\
Rochester Institute of Technology\\
Rochester, NY 14623 \\
\and
{Lane A. Hemaspaandra}\\
Department of Computer Science\\ University of Rochester\\
Rochester, NY 14627\\
\and
{J\"org Rothe}\thanks{Work done in part 
while visiting the University of Rochester.}
\\
Institut f\"ur Informatik\\
Heinrich-Heine-Universit\"at D\"usseldorf\\
40225 D\"usseldorf, Germany  \\
}

\date{December 9, 2007}

\maketitle
\begin{abstract}
Control and bribery are settings in which an external agent seeks to
influence the outcome of an election.  Faliszewski et
al.~\cite{fal-hem-hem-rot:c-ABBREVIATE:llull} proved that 
Llull
voting
(which is 
here denoted by Copeland$^1$) 
and a variant 
(here denoted by Copeland$^0$) 
of
Copeland voting 
are computationally resistant
to many, yet not all, types of constructive control and that they also
provide broad resistance to bribery.  We study a parameterized version
of Copeland voting, denoted by $\copelandalpha$, where the parameter
$\alpha$ is a rational number between $0$ and $1$ that specifies how
ties are
valued
in the pairwise comparisons of candidates in Copeland
elections.  We establish resistance or vulnerability results, in every
previously studied control scenario, for $\copelandalpha$ for each
rational $\alpha$, $0 < \alpha < 1$.  In particular, we prove that
Copeland$^{0.5}$, the system commonly referred to as ``Copeland
voting,'' provides full resistance to constructive control.
Among the systems with a polynomial-time winner problem, this is the
first natural election system proven to have 
full resistance to constructive control.
Results on bribery and
fixed-parameter tractability of bounded-case control 
proven
for Copeland$^0$ and Copeland$^1$
in~\cite{fal-hem-hem-rot:c-ABBREVIATE:llull} are extended to
$\copelandalpha$ for each rational $\alpha$, $0 < \alpha < 1$;
we also give results in more flexible models such 
as microbribery and extended control.
\end{abstract}

\section{Introduction}

Preference aggregation by voting procedures has 
been the focus of much
attention within the field of multiagent systems.
Agents (called voters in the context of voting) may have different,
often conflicting individual preferences over the given alternatives
(or candidates).  Voting rules (or, synonymously, election systems)
provide a useful method for them to come to a ``reasonable'' 
decision
on which alternative to choose.  One key issue here is that there
might be attempts to influence the outcome of elections.
Settings in which such influence on elections
can be implemented include
manipulation~\cite{con-san-lan:j:when-hard-to-manipulate,pro-ros-zoh:c-ABBREVIATE:multiwinner},
electoral
control~\cite{bar-tov-tri:j:control,fal-hem-hem-rot:c-ABBREVIATE:llull,hem-hem-rot:j:destructive-control,hem-hem-rot:c-ABBREVIATE:hybrid,pro-ros-zoh:c-ABBREVIATE:multiwinner},
and
bribery~\cite{fal-hem-hem:c-ABBREVIATE:bribery,fal-hem-hem-rot:c-ABBREVIATE:llull}.
Although 
reasonable election systems typically are susceptible
to these kinds of influence (for manipulation
this is universally true, via 
the Gibbard--Satterthwaite and Duggan--Schwartz Theorems),
computational complexity can be used to
provide some protection in each such setting.  We study the extent to
which the Copeland election system~\cite{cop:m:copeland} (see also
\cite{saa-mer:j:copeland1,mer-saa:j:copeland2}; a similar system
was also studied by Zermelo) resists,
computationally, control and bribery attempts.

Copeland elections are one of the classical voting procedures that are
based on pairwise comparisons of candidates: The winner (by a strict
majority of votes) of each such a head-to-head contest is 
awarded 
one
point and the loser receives no point; whoever collects the most
points over all these contests (including tie-related points)
is the election's winner.  The
points awarded 
for ties in such head-to-head majority-rule contests are treated
in various ways in the literature.  Faliszewski et
al.~\cite{fal-hem-hem-rot:c-ABBREVIATE:llull} proposed a parameterized
version of Copeland elections, denoted by $\copelandalpha$, where the
parameter $\alpha$ is a rational number between $0$ and $1$ such that,
in case of a tie, both candidates receive $\alpha$ points.  
So 
the system widely referred to in the literature as
``Copeland elections'' is Copeland$^{0.5}$, where tied candidates
receive half a point each (see, e.g., Merlin and
Saari~\cite{saa-mer:j:copeland1,mer-saa:j:copeland2}; the definition
used by Conitzer et al.~\cite{con-san-lan:j:when-hard-to-manipulate}
can be scaled to be equivalent to Copeland$^{0.5}$).
Copeland$^0$, where
tied candidates come away empty-handed, has sometimes 
also been referred to 
as
``Copeland elections''
(see, e.g.,
\cite{pro-ros-kam:c-ABBREVIATE:noisy,fal-hem-hem-rot:c-ABBREVIATE:llull}).
An
election 
system proposed by the 
Catalan 
philosopher 
and
theologian 
Ramon Llull in the 13th century (see, e.g., the references
in~\cite{fal-hem-hem-rot:c-ABBREVIATE:llull}) is in this notation
nothing other than Copeland$^1$, where tied candidates are awarded one
point each, just like winners of head-to-head contests.

Faliszewski et al.~\cite{fal-hem-hem-rot:c-ABBREVIATE:llull} studied
the systems Copeland$^0$ and Copeland$^1$ with respect to their
(computational) resistance and vulnerability to bribery and procedural
control.  Bribery and control are settings in which an external actor
seeks to influence the outcome of an election.  Bribery is somewhat
akin to electoral manipulation and strategic voting in that the briber
tries to reach his or her goal by bribing some voters 
to
change their preferences.  (The difference between bribery and
manipulation is that manipulative voters themselves cast their votes
insincerely, i.e., there is no external agent.)  In contrast, the
external actor in control scenarios (who by 
tradition is, potentially confusingly, called ``the chair'')
seeks to reach this
goal via changing the election procedure, namely via
adding/deleting/partitioning either candidates or voters.

Bartholdi, Tovey, and Trick~\cite{bar-tov-tri:j:control} were the
first to study the computational aspects of control: How hard is it,
computationally, for the chair to exert control?  In their seminal
paper they introduced a number of fundamental control scenarios 
involving
(what is now called) \emph{constructive} control, i.e.,
where the chair's goal is to
make some designated candidate win.  Other papers studying control
include~\cite{hem-hem-rot:j:destructive-control,pro-ros-zoh:c-ABBREVIATE:multiwinner,hem-hem-rot:c-ABBREVIATE:hybrid,fal-hem-hem-rot:c-ABBREVIATE:llull},
which in addition to constructive control also consider
\emph{destructive} control, where the chair tries to preclude some
designated
candidate from winning.  The notion of bribery in elections
was introduced by Faliszewski et
al.~\cite{fal-hem-hem:c-ABBREVIATE:bribery} and was also studied
in~\cite{fal-hem-hem-rot:c-ABBREVIATE:llull}.

At first glance, one might be tempted to think that the definitional
perturbation due to the parameter $\alpha$ in $\copelandalpha$
elections is negligible.  However, as noted
in~\cite{fal-hem-hem-rot:c-ABBREVIATE:llull}, ``\ldots\ it can make
the dynamics of Llull's system quite different from those of
[Copeland$^0$].  Proofs of results for Llull differ considerably from
those for [Copeland$^0$].''  This statement notwithstanding, we show that in
most cases it is possible to obtain a unified---though sometimes
rather involved---construction that works for both systems, and even
for $\copelandalpha$ with respect to every rational~$\alpha$, $0 \leq
\alpha \leq 1$.  In particular,
we establish resistance or vulnerability results 
for Copeland$^{0.5}$ 
(which is the
system commonly referred to as ``Copeland'')
in every
previously studied control scenario.\footnote{Also, 
  our new results apply, e.g., to  events such as the group stage of
  FIFA world-cup finals, which is, in essence, a
  series of $\copelandalpha$ tournaments with $\alpha = \frac{1}{3}$.}
In doing so, we provide
an example of a control problem where the complexity of
Copeland$^{0.5}$ differs from that of both Copeland$^0$ and
Copeland$^1$: While the latter two problems are vulnerable to
constructive control by adding (an unlimited number of) candidates,
Copeland$^{0.5}$ is resistant to this control type (see
Section~\ref{sec:prelims} for definitions and Theorem~\ref{thm:ccacu}
for this result).

Thus Copeland (i.e., Copeland$^{0.5}$) 
is the first natural election system with a
polynomial-time winner problem that is proven to be resistant to every
type of constructive control that has been proposed in the literature
to date.  Moreover, if one uses the hybridization method of
Hemaspaandra et al.~\cite{hem-hem-rot:c-ABBREVIATE:hybrid} to combine
this full resistance of Copeland$^{0.5}$ to constructive control with
the full resistance of Copeland$^{0.5}$ to destructive voter control
(which we also prove here) and with the full resistance of
plurality\footnote{In plurality-rule elections, every voter gives one
point to his or her most preferred candidate.  Whoever collects the
most points is this election's plurality winner.}
to destructive candidate control
(see~\cite{hem-hem-rot:j:destructive-control,fal-hem-hem-rot:c-ABBREVIATE:llull}),
one obtains a hybrid election system that (a)~is resistant to every
(constructive and destructive)
control type previously considered in the literature, (b)~has a
polynomial-time winner problem, and (c)~has only natural election
systems as its constituents.  In contrast, one of the constituent systems
for the hybrid constructed
in~\cite{hem-hem-rot:c-ABBREVIATE:hybrid}, which is there shown to
resist twenty control types, is rather artificial.

\section{Preliminaries}
\label{sec:prelims}

An election is specified by a finite set $C$ of candidates and a
finite collection $V$ of voters, where each voter has preferences over
the candidates.  We consider both rational and irrational voters.  The
preferences of a rational voter are expressed by a preference list of
the form $a > b > c$ (assuming $C = \{a, b, c\}$), where the
underlying relation $>$ is a strict linear order that is transitive.
The preferences of an irrational voter are expressed by a preference
table that, for any two distinct candidates, specifies which of them
is preferred to the other by this voter.  An election system is a rule
that determines the winner(s) of each given election $(C,V)$.  In this
paper, we consider a parameterized version of Copeland's election
system~\cite{cop:m:copeland}, denoted $\copelandalpha$, where the
parameter $\alpha$ is a rational number between $0$ and $1$ that
specifies how ties are rewarded in the head-to-head majority-rule
contests between any two distinct candidates.

\begin{definition}[\cite{fal-hem-hem-rot:c-ABBREVIATE:llull}]
\label{def:copeland}
  Let $\alpha$, $0 \leq \alpha \leq 1$, be a fixed rational number.
  In a $\copelandalpha$ election, the voters indicate which among any
  two distinct candidates they prefer.  For each such head-to-head
  contest, if some candidate is preferred by a strict majority of
  voters then he or she obtains one point and the other candidate
  obtains zero points, and if a tie occurs then both candidates obtain
  $\alpha$ points.
  Let $E = (C,V)$ be an election.  For each $c \in C$,
  $\copelandalphascore_{E}(c)$ is the sum of $c$'s $\copelandalpha$
  points in~$E$.  Every candidate $c$ with maximum
  $\copelandalphascore_{E}(c)$ wins.

  Let $\copelandalphairrational$ denote the same election system but
  with voters allowed to be irrational.
\end{definition}

In the literature, the term ``Copeland elections'' is most often used
for the system Copeland$^{0.5}$, and is sometimes used for
Copeland$^0$.  The system Copeland$^1$ was proposed by Llull already
in the 13th century (see the references
in~\cite{fal-hem-hem-rot:c-ABBREVIATE:llull}) and so is called Llull
voting.

We now define the control problems we consider, in both the
constructive and the destructive version.
Let $\electionsystem$ be an election system. In our case,
$\electionsystem$ will be either $\copelandalpha$ or
$\copelandalphairrational$, where $\alpha$, $0 \leq \alpha \leq 1$, is
a fixed rational number.  In fact, since the types of control we
consider here are well-known from the literature (see, e.g.,
\cite{bar-tov-tri:j:control,fal-hem-hem-rot:c-ABBREVIATE:llull,hem-hem-rot:j:destructive-control}),
we will content ourselves with the definition of some examples of
these problems (in particular some of those that occur in the proofs
to be presented in Section~\ref{sec:control-candidate} below).

We start with defining control via adding candidates.  Note that there
are two versions of this control type.  The \emph{unlimited} version
(which, for the constructive case, was introduced by Bartholdi, Tovey,
and Trick~\cite{bar-tov-tri:j:control}) asks whether the election
chair can add (any number of) candidates from
a given
pool of spoiler candidates
in order to either make his or her favorite
candidate win the election (in the constructive case), or prevent his or her
despised candidate from winning (in the destructive case):

\begin{description}
\item[Name:] $\scontrol{\electionsystem}{CCAC$_{\rm u}$}$ and
  $\scontrol{\electionsystem}{DCAC$_{\rm u}$}$.
\item[Given:] Disjoint candidate sets $C$ and $D$, a collection $V$ of
voters represented via their preference lists (or preference tables in
the irrational case) over the candidates in $C \cup D$, and a
distinguished candidate~$p \in C$.
\item[Question ($\scontrol{\electionsystem}{CCAC$_{\rm u}$}$):] Does
there exist a subset $D'$ of $D$ such that $p$ is a winner of the
$\electionsystem$ election with candidates $C \cup D'$ and voters $V$?
\item[Question ($\scontrol{\electionsystem}{DCAC$_{\rm u}$}$):] Does
there exist a subset $D'$ of $D$ such that $p$ is not a winner of the
$\electionsystem$ election with candidates $C \cup D'$ and voters $V$?
\end{description}

The only difference in the \emph{limited} version of constructive and
destructive control via adding candidates
($\scontrol{\electionsystem}{CCAC}$ and
$\scontrol{\electionsystem}{DCAC}$, for short) is that the chair needs
to achieve his or her goal by adding at most $k$ candidates from the
given set of spoiler candidates.  This version of control by adding
candidates was proposed in~\cite{fal-hem-hem-rot:c-ABBREVIATE:llull}
to synchronize the definition of control by adding candidates with the
definitions of control by deleting candidates, adding voters, and
deleting voters.

Our second example regards control via run-off partition of
candidates, where we focus on the constructive case:

\begin{description}
\item[Name:] $\scontrol{\electionsystem}{CCRPC-TP}$ (respectively,
  $\scontrol{\electionsystem}{CCRPC-TE}$).
\item[Given:] A set $C$ of candidates and a collection $V$ of voters
represented via their preference lists (or preference tables in the
irrational case) over~$C$, a distinguished candidate~$p \in C$, and a
nonnegative integer~$k$.
\item[Question:] Is it possible to partition $C$ into $C_1$ and $C_2$
  such that $p$ is a winner of the two-stage election where the
  winners of subelection $(C_1,V)$ that survive the tie-handling rule
  (TP or TE) compete against the winners of subelection $(C_2,V)$ that
  survive the tie-handling rule?  (Subelections are conducted using
  system~$\electionsystem$.)
\end{description}

As one can see from the above examples, we use the following naming
conventions for control problems.  The name of a control problem
starts with the election system used (when clear from context, it may
be dropped), followed by CC for ``constructive control'' or by DC for
``destructive control,'' followed by the acronym of the type of
control: AC for ``adding (a limited number of) candidates,'' AC$_{\rm
u}$ for ``adding (an unlimited number of) candidates,'' DC for
``deleting candidates,'' PC for ``partition of candidates,'' RPC for
``run-off partition of candidates,'' AV for ``adding voters,'' DV for
``deleting voters,'' and PV for ``partition of voters,'' and all the
partitioning cases (PC, RPC, and PV) are followed by the acronym of
the tie-handling rule used in subelections, namely TP for ``ties
promote'' (i.e., all winners of a given subelection are promoted to
the final round of the election) and TE for ``ties eliminate'' (i.e.,
if there is more than one winner in a given subelection then none of
this subelection's winners is promoted to the final round of the
election).

We now turn to the definition of bribery problems
(see~\cite{fal-hem-hem:c-ABBREVIATE:bribery}), where the briber seeks
to reach his or her goal via bribing certain voters to make them
change their preferences.

\begin{description}
\item[Name:] $\sbribery{\electionsystem}$.
\item[Given:] A set $C$ of candidates, a collection $V$ of voters
  represented via their preference lists (or preference tables in the
  irrational case) over~$C$, a distinguished candidate $p \in C$, and
  a nonnegative integer~$k$.
\item[Question:] Does there exist a voter collection $V'$ over~$C$,
where $V'$ results from $V$ by modifying at most $k$ voters, such that
$p$ wins the $\electionsystem$ election $(C,V')$?
\end{description}

For $\sdestbribery{\electionsystem}$, the destructive bribery problem
for $\electionsystem$, we require $p$ to be \emph{not} a winner.

Note that the above definitions focus on \emph{a winner}, i.e., they
are in the \emph{nonunique-winner model}.  The \emph{unique-winner}
analogs of these problems can be defined by requiring the
distinguished candidate $p$ to be the unique winner (or to not be a
unique winner in the destructive case).
Let $\electionsystem$ be an election system and let $\Phi$ be a
control type.  We say $\electionsystem$ is \emph{immune to
$\Phi$-control} if the chair can never reach his or her goal (of
making a given candidate win in the constructive case, and of blocking
a given candidate from winning in the destructive case) via asserting
$\Phi$-control.  $\electionsystem$ is said to be \emph{susceptible to
$\Phi$-control} if $\electionsystem$ is not immune to $\Phi$-control.
$\electionsystem$ is said to be \emph{vulnerable to $\Phi$-control} if
it is susceptible to $\Phi$-control and there is a polynomial-time
algorithm for solving the control problem associated with~$\Phi$.
$\electionsystem$ is said to be \emph{resistant to $\Phi$-control} if
it is susceptible to $\Phi$-control and the control problem associated
with $\Phi$ is $\np$-hard.  The above notions were introduced by
Bartholdi, Tovey, and Trick~\cite{bar-tov-tri:j:control}
(see also, e.g.,
\cite{hem-hem-rot:j:destructive-control,pro-ros-zoh:c-ABBREVIATE:multiwinner,hem-hem-rot:c-ABBREVIATE:hybrid,fal-hem-hem-rot:c-ABBREVIATE:llull}).
We say $\electionsystem$ is \emph{vulnerable to constructive
(respectively, destructive) bribery} if $\sbribery{\electionsystem}$
(respectively, $\sdestbribery{\electionsystem}$) is in~$\p$.  We say
$\electionsystem$ is \emph{resistant to constructive (respectively,
destructive) bribery} if $\sbribery{\electionsystem}$ (respectively,
$\sdestbribery{\electionsystem}$) is $\np$-hard.

Many of our reductions in Section~\ref{sec:control} are from the
$\np$-complete vertex cover problem: Given an undirected graph $G =
(V(G),E(G))$ and a nonnegative integer~$k$, does there exist a set $W$
such that $W \subseteq V(G)$, $\|W\| \leq k$, and for every edge $e =
\{ u, v \}$, $e \in E(G)$, it holds that $e \cap W \neq \emptyset$?

The study of fixed-parameter complexity 
(see, e.g.,~\cite{dow-fel:b:parameterized})
has been expanding explosively
since it was parented as a field by Downey, Fellows, and others in the
late 1980s and the 1990s.  Although the area has built a rich variety
of complexity classes regarding parameterized problems, for the
purpose of the current paper we need focus only on one very
important class, namely, the
class $\fpt$.  Briefly put, a problem parameterized by some value $j$
(which, note, can be viewed as a family of problems, one per value of
$j$) is said to be \emph{fixed-parameter tractable} (equivalently, to
belong to the class $\fpt$) 
if there is an algorithm for the problem whose running time is 
$f(k)n^{O(1)}$.

In our context, we
consider two parameterizations: bounding the
number of candidates and bounding the number of voters.  We 
use the same notations used throughout this paper to describe
problems, except we
postpend a ``-BV${}_j$'' to a problem name to
state that the number of voters may be at most $j$, and we
postpend a ``-BC${}_j$'' to a problem name to state that the number of
candidates may be at most $j$.  In each case, the bound applies to the
full number of such items involved in the problem.  For example, in
the case of control by adding voters, the $j$ must bound the total of
the number of voters in the election added together with the number of
voters in the pool of voters available for adding.

\section{Control}
\label{sec:control}

\subsection{Overview of Results}
\label{sec:overview}

Our main result regarding control is Theorem~\ref{thm:control} below.

\begin{theorem}
  \label{thm:control}
  For each rational $\alpha$, $0 \leq \alpha \leq 1$, $\copelandalpha$
  elections are resistant and vulnerable to control as shown in
  Table~\ref{tab:control}, both for rational and irrational voters and
  in both the nonunique-winner model and the unique-winner model.
\end{theorem}

\begin{table}[t]
\centering
\begin{tabular}{|l|c|c|c|c|c|c|}
\hline 
\multicolumn{1}{|c|}{}
 & \multicolumn{6}{|c|}{$\copelandalpha$}
 \\ \cline{2-7}
\multicolumn{1}{|c|}{}
 & \multicolumn{2}{c|}{$\alpha = 0$}
 & \multicolumn{2}{c|}{$0 < \alpha < 1$}
 & \multicolumn{2}{c|}{$\alpha = 1$}
 \\
\hline 
Control type  & CC & DC & CC & DC & CC & DC \\
\hline
AC$_{\rm u}$  & V  & V  & {\bf R}  & {\bf V}  & V  & V  \\
AC            & R  & V  & {\bf R}  & {\bf V}  & R  & V  \\
DC            & R  & V  & {\bf R}  & {\bf V}  & R  & V  \\
RPC-TP        & R  & V  & {\bf R}  & {\bf V}  & R  & V  \\
RPC-TE        & R  & V  & {\bf R}  & {\bf V}  & R  & V  \\
PC-TP         & R  & V  & {\bf R}  & {\bf V}  & R  & V  \\
PC-TE         & R  & V  & {\bf R}  & {\bf V}  & R  & V  \\
\hline
PV-TE         & R  & R  & {\bf R}  & {\bf R}  & R  & R  \\
PV-TP         & R  & R  & {\bf R}  & {\bf R}  & R  & R  \\
AV            & R  & R  & {\bf R}  & {\bf R}  & R  & R  \\
DV            & R  & R  & {\bf R}  & {\bf R}  & R  & R  \\
\hline
\end{tabular}
\caption{\label{tab:control} Resistance (R) and vulnerability (V) of
{\boldmath Copeland$^\alpha$ elections, for rationals $\alpha$, $0 \leq
\alpha \leq 1$}.}
\end{table}

Boldface results in Table~\ref{tab:control} are new to this paper and
nonboldface results are due to Faliszewski et
al.~\cite{fal-hem-hem-rot:c-ABBREVIATE:llull}.  Note that the notion
widely referred to in the literature simply as ``Copeland elections,''
which we here for clarity call Copeland$^{0.5}$, possesses all ten of
our basic types
of constructive resistance and, in addition, even has constructive
AC$_{\rm u}$ resistance.  These resistances should be compared with
the results known for the other notion that in the literature is
occasionally referred to as ``Copeland elections,'' namely
Copeland$^0$, and with the results known for Llull elections, which
are here denoted by Copeland$^1$,
see~\cite{fal-hem-hem-rot:c-ABBREVIATE:llull}.  While Copeland$^0$ and
Copeland$^1$ possess all ten of our basic types of constructive
resistance, they both are vulnerable to this eleventh type of
constructive control, the incongruous but historically resonant notion
of constructive control by adding an unlimited number of candidates
(i.e., CCAC$_{\rm u}$).

It is known that plurality is resistant to the six basic types of
destructive candidate control and also to DCAC$_{\rm u}$,
see~\cite{hem-hem-rot:j:destructive-control,fal-hem-hem-rot:c-ABBREVIATE:llull}.
Since by Theorem~\ref{thm:control}, Copeland$^{0.5}$ provides
resistance for all ten basic constructive control types and for
CCAC$_{\rm u}$, and also for the four basic types of destructive voter
control, the hybrid (in the sense
of~\cite{hem-hem-rot:c-ABBREVIATE:hybrid}) of plurality with
Copeland$^{0.5}$ is resistant to each basic type of constructive and
destructive control and in addition to constructive and destructive
AC$_{\rm u}$ control.  This result follows via
Theorem~\ref{thm:control} and the results of Hemaspaandra et
al.~\cite{hem-hem-rot:c-ABBREVIATE:hybrid}.  And, unlike the hybrid
system constructed by Hemaspaandra et
al.~\cite{hem-hem-rot:c-ABBREVIATE:hybrid}, this hybrid uses only
natural systems as its constituents.

\begin{corollary}
\label{cor:control}
The hybrid (in the sense of~\cite{hem-hem-rot:c-ABBREVIATE:hybrid}) of
plurality and Copeland$^{0.5}$ is resistant to each of the twenty
basic types of constructive and destructive control and also to
constructive and destructive AC$_{\rm u}$ control, and it has a
polynomial-time winner problem.
\end{corollary}

The next two sections discuss the single results contained in
Theorem~\ref{thm:control} in more detail and sketch some of the
proofs.  All the results stated in
Sections~\ref{sec:control-candidate} and~\ref{sec:control-voter}
are true both in the rational and irrational voter model and in both
the nonunique-winner model and the unique-winner model.

\subsection{Candidate Control}
\label{sec:control-candidate}

We start with candidate control.  Faliszewski et
al.~\cite{fal-hem-hem-rot:c-ABBREVIATE:llull} showed that both
Copeland$^0$ and Copeland$^1$ are vulnerable to each destructive
control type in Table~\ref{tab:control}.  To extend these results to
$\copelandalpha$ elections for each rational $\alpha$, $0 \leq \alpha
\leq 1$, our proofs for destructive control by adding and deleting
candidates use the following observation.
Let $(C,V)$ be an election, and let $\alpha$ be a fixed rational
number such that $0 \leq \alpha \leq 1$.
For every candidate $c \in C$ it holds that:
$\copelandalphascore_{(C,V)}(c) = \sum_{d \in C - \{c\}}
\copelandalphascore_{(\{c,d\},V)}(c).$

The candidate partition and run-off partition cases can be shown to
reduce to the case of deleting candidates, and the vulnerability results
in Theorem~\ref{thm:control-candidate-destructive}
use greedy algorithms.

\begin{theorem}
\label{thm:control-candidate-destructive}
 For each rational number $\alpha$, $0 \leq \alpha \leq 1$,
 $\copelandalpha$ is vulnerable to destructive control via
(a)~adding candidates (both DCAC and DCAC$_{\rm u}$, i.e., both for
 a limited and an unlimited number of candidates),
(b)~deleting candidates (DCDC),
(c)~partition of candidates (in both the TP and TE model, i.e.,
  DCPC-TP and DCPC-TE), and
(d)~run-off partition of candidates (in both the TP and TE model,
i.e., DCRPC-TP and DCRPC-TE),
\end{theorem}

Turning now to constructive candidate control, our resistance proofs
use the following two lemmas, which we here state without proof.
Lemma~\ref{thm:pad-election} shows how to construct a ``padded''
election with useful properties.  Lemma~\ref{thm:construction-lemma}
then shows how to build an election via combining smaller ones.

\begin{lemma}
\label{thm:pad-election}
  Let $\alpha$ be a rational number such that $0 \leq \alpha \leq 1$.
  For each positive integer~$n$, there is an election $\pad_n = (C,V)$
  such that $\|C\| = 2n+1$ and, for each candidate $c \in C$, it holds
  that $\copelandalphascore_{\pad_n}(c) = n$.
\end{lemma}

\begin{lemma}
  \label{thm:construction-lemma}
  Let $E = (C,V)$ be an election where $C = \{c_1, \ldots, c_n\}$, and
  let $\alpha$ be a rational number such that $0 \leq \alpha \leq 1$.  For each
  candidate~$c_i$, we denote the number of head-to-head ties of $c_i$ in
  $E$ by~$t_i$.
  Let $k_1, \ldots , k_n$
  be a sequence of $n$ nonnegative integers such that for each $k_i$ we
  have $0 \leq k_i \leq n$. There is an election
  $E' = (C',V')$ such that:
(a)~$C' = C \cup D$, where $D = \{d_1, \ldots, d_{2n^2}\}$;
(b)~for each $i$, $1 \leq i \leq n$,
   $\copelandalphascore_{E'}(c_i) = 2n^2 - k_i + \alpha t_i$;
(c)~for each $i$, $1 \leq i \leq 2n^2$,
   $\copelandalphascore_{E'}(d_{i}) \leq n^2+1$.
\end{lemma}

Faliszewski et al.~\cite{fal-hem-hem-rot:c-ABBREVIATE:llull} showed
that both Copeland$^0$ and Copeland$^1$ are resistant to constructive
control via adding (a limited number of) candidates.  This is subsumed
by the following more general result.

\begin{theorem}
  \label{thm:ccac}
  For each rational number $\alpha$ such that $0 \leq \alpha \leq 1$,
  $\copelandalpha$ is resistant to constructive control via adding
  candidates (CCAC).
\end{theorem}

In contrast with the known result that both Copeland$^0$ and
Copeland$^1$ are vulnerable to constructive control via adding an
unlimited number of
candidates~\cite{fal-hem-hem-rot:c-ABBREVIATE:llull}, we show that
$\copelandalpha$ is resistant to this control type if $0 < \alpha <
1$.

Notation: In the proofs of Section~\ref{sec:control-candidate}, we
often identify an election with its set of candidates, since in the
case of candidate control the set of voters is fixed and cannot change.

\begin{theorem}
  \label{thm:ccacu}
  For each rational number $\alpha$, $0 < \alpha < 1$,
  $\copelandalpha$ is resistant to constructive control via adding an
  unlimited number of candidates (CCAC$_{\mathrm u}$).
\end{theorem}

\begin{proofs}
We provide a reduction from the vertex cover problem.
Let $(G,k)$ be an instance of the vertex cover problem, where $G$ is
an undirected graph and $k$ is the bound on the size of the vertex
cover that we seek.  Let $E(G) = \{e_1, \ldots,e_m\}$ be the set of
$G$'s edges and $V(G) = \{1, \ldots, n\}$ be the set of $G$'s
vertices. Using Lemma~\ref{thm:construction-lemma}, we can build an
election $E' = (C,V')$
such that:
(a) $\|C\| = 2\ell^2 + \ell$, where $\ell = 2n+2m$;
(b) $\{p, r, e_1, \ldots, e_m\} \subseteq C$ (the remaining
    candidates are used for padding);
(c) $\copelandalphascore_{E'}(p) = 2\ell^2 -2 $;
(d) $\copelandalphascore_{E'}(r) = 2\ell^2 -2 -k + k\alpha$
    in the nonunique-winner case (respectively,
    $\copelandalphascore_{E'}(r) = 2\ell^2 -2 -k + (k-1)\alpha$
    in the unique-winner case);
(e) for each $e_i \in C$,
    $\copelandalphascore_{E'}(e_i) = 2\ell^2 -2 + \alpha$
    in the non\-unique-winner case (respectively,
    $\copelandalphascore_{E'}(e_i) = 2\ell^2 - 2$ in the unique-winner case);
(f) the scores of all candidates other than $p, e_1, \ldots, e_m$
    are at most $2\ell^2-n-2$.

(We omit the details of the construction, but mention that one can
start from
any election with at least $2n+2m$ candidates,
add some ties between some padding candidates and the $e_i$'s if needed, and
then apply Lemma~\ref{thm:construction-lemma}.)

We form election $E = (C \cup D,V)$ from $E'$ via adding candidates 
$D = \{1, \ldots, n\}$
and appropriate voters such that the results of head-to-head contests are:
(1) $p$ ties with all candidates in $D$;
(2) for each $e_j$, if $e_j$ is incident with $i \in D$ then candidate
    $i$ defeats candidate $e_j$,
 and otherwise they tie;
(3) all other candidates in $C'$ defeat each of the candidates in~$D$.

Our instance of CCAC$_\mathrm{u}$ is formed by the candidate set $C$
(the candidates already enrolled in the election), the candidate set
$D$ (the candidates that can be added to the election), and the set of
voters $V$, where each voter has preferences over the candidates in $C
\cup D$. Note that the candidates in $D$ correspond to the vertices of~$G$.
We claim that there is a set $D'$ (where $D' \subseteq D$) such that $p$ is
a winner (respectively, the unique winner) of the
$\copelandalpha$ election $(C \cup D',V)$
if and only if $G$ has a vertex cover of size at most~$k$.

It is easy to see that if $D'$ corresponds to a vertex cover of size
at most $k$ then $p$ is a winner (respectively, the unique winner) of
$\copelandalpha$ election $C \cup D'$. The reason is that adding any one
member of $D'$ increases $p$'s score by $\alpha$, increases $r$'s
score by one, and for each $e_j$, adding $i \in D'$ increases $e_j$'s
score by $\alpha$ if and only if $e_j$ is not incident with $i$. Thus,
the nonpadding candidates in $C \cup D'$ have the following scores
in the resulting election $E''$ with candidates $C \cup D'$ (it
is clear that none of the padding candidates has enough
$\copelandalpha$ points to become a winner after adding any subset of
candidates from~$D$):
(a) $\copelandalphascore_{E''}(p) = 2\ell^2 -2 + k\alpha$;
(b) $\copelandalphascore_{E''}(r) = 2\ell^2 -2 + k\alpha$
   in the nonunique-winner case (respectively,
   $2\ell^2 -2 + (k-1)\alpha$ in the unique-winner case);
(c) $\copelandalphascore_{E''}(e_i) \leq 2\ell^2 -2 + k\alpha$
   in the nonunique-winner case (respectively,
   $2\ell^2 -2 + (k-1)\alpha$ in the unique-winner case).

As a result, we see that adding all members of $D'$ makes $p$ a winner
(respectively, the unique winner).

On the other hand, assume that $p$ can become a winner via adding some
subset $D'$ of candidates from $D$. First, note that $\|D'\| \leq k$,
since otherwise $r$ would end up with more points (respectively, 
at least as many
points) as $p$ and so $p$ would not be a winner (respectively, 
would not be the unique winner).  We
claim that $D'$ corresponds to a vertex cover of $G$. For the sake of
contradiction assume that there is some edge $e_j$ incident to
vertices $u$ and $v$ such that neither $u$ nor $v$ is in $D'$.
However, if this
were the case then candidate $e_j$ would have more
points (respectively, 
at least as many points) as $p$ and so $p$ would not be a
winner (respectively, 
would not be the unique winner).  Thus, $D'$ must form a vertex cover of
size at most~$k$.~\end{proofs}

The following result extends to all rationals $\alpha$, $0 \leq \alpha
\leq 1$, the known result that Copeland$^0$ and Copeland$^1$ are
resistant to constructive control via deleting
candidates~\cite{fal-hem-hem-rot:c-ABBREVIATE:llull}.

\begin{theorem}
  \label{thm:ccdc}
  Let $\alpha$ be a rational number such that $0 \leq \alpha \leq 1$.
  $\copelandalpha$
  is resistant to constructive control via deleting candidates (CCDC).
\end{theorem}

\begin{proofs}
The proof follows
via a reduction from the vertex cover problem.  We first handle the
nonunique-winner case.

Let $(G,k)$ be the input instance of the vertex cover problem, where
$G$ is an undirected graph and $k$ is the upper bound on the size of
the vertex cover that we seek. Let $V(G) = \{1, \ldots, n\}$ and let
$E(G) = \{e_1, \ldots, e_m\}$.  We build election $E' = (C',V')$,
where $C' = \{p, r, e_1, \ldots, e_m, 1, \ldots, n\}$ and the voter
set $V'$ yields the following results of head-to-head contests
(omitting the details of the construction due to space):
(1) $p$ defeats $r$;
(2) each candidate $e_i \in C$ defeats exactly those two
    candidates $u, v \in \{1, \ldots, n\}$ that the edge $e_i$ is
    incident with;
(3) each candidate $u \in \{1, \ldots, n\}$ defeats both $p$ and
    all candidates $e_i$ such that vertex $u$ is not incident to
    $e_i$;
(4) all the other contests result in a tie.

Let $\ell = n+m$. We form an election $E = (C,V)$ via combining
election $E'$ with election $\pad_\ell = (C'',V'')$, where $C'' =
\{t_0, \ldots, t_{2\ell}\}$ and the set $V''$ of voters is set as in
Lemma~\ref{thm:pad-election}.  We select the following results of
head-to-head contest between the candidates in $C'$ and the candidates
in $C''$: $p$ and all candidates $e_i \in C$ defeat everyone in $C''$
and each candidate in $C''$ defeats all candidates in $C' - \{p, e_1,
\ldots , e_m\}$.
It is easy to verify that election $E$ yields the following $\copelandalpha$
scores:
(a) $\copelandalphascore_E(p) = m\alpha + 1+ 2\ell+1$;
(b) $\copelandalphascore_E(r) = m + n\alpha$;
(c) for each $e_i \in C$, $\copelandalphascore_E(e_i) = m\alpha + 2+ 2\ell+1$;
(d) for each $i \in C$, $\copelandalphascore_E(i) \leq 1 + m +n\alpha$;
(e) for each $t_i \in C$, $\copelandalphascore_E(t_i) = \ell + n + 1$.

Thus, the set of winners of $E$ is $W = \{e_1, \ldots, e_m\}$.
We claim that $p$ can become a winner of $\copelandalpha$ election $E$
via deleting at most $k$ candidates if and only if the graph
$G$ has a vertex cover of size at most $k$.

First note that if $k \geq n$ then $G$ obviously has a vertex
cover of size at most $k$ (namely, the set of all the vertices) and so
from now on we assume $k < n$.  Also, it is easy to see that if $k
\geq m$ (i.e., if our vertex cover can have more elements than there
are edges) then clearly a vertex cover exists and so we assume that $k
< m$.  Also, we note that all candidates except $p$ lose by at
least $n+1$ points to each of the winners and so $p$ is the only
candidate that can possibly become a winner via deleting at most $k
\leq n$ candidates.

Assume that $p$ can become a winner via deleting at most $k <
n$ candidates, and let $D \subseteq C$ be a smallest set such that
deleting exactly the candidates in $D$ from election $E$ guarantees
$p$'s victory. We start by observing that $D$ necessarily contains
only candidates that correspond to vertices of~$G$.  For the sake
of contradiction, assume that $D$ does contain some candidate $d$ such
that $d \notin \{1, \ldots, n\}$.  Clearly, $d$ cannot be $r$,
since deleting $r$
decreases $p$'s score without changing the score of any of the
candidates in $W$ and so removing $r$ from $D$ would yield a smaller
set of candidates whose deletion guarantees $p$'s victory. Similarly,
$d$ cannot be any other non-vertex candidate, since deleting $d$ from
election $E$ would affect the score of $p$ and the scores of all
remaining candidates from $\{e_1, \ldots e_m\}$
in the same way.\footnote{Keep in mind
  that $d$ could be one of the candidates $e_1$ through $e_m$.
  However, since $k < m$, $D$ cannot contain all of these
  candidates.}
Thus, again, removing $d$ from $D$
would yield a smaller set with the required property.

Now note that, in election $E$, each of $e_1, \ldots, e_m$ has exactly
one $\copelandalpha$ point of advantage over $p$. Deleting any
candidate $u$ corresponding to a vertex of $G$ does not affect $p$'s
score but it does lower by one the score of all the candidates $e_1,
\ldots, e_m$ that correspond to the edges incident with $u$. Since
deleting the candidates in $D$ makes $p$ a winner and since $D$ contains
only up to $k$ candidates that correspond to vertices in $G$, it must
be the case that the candidates in $D$ correspond to a vertex cover of $G$
having size at most~$k$.

For the converse, it is easy to see that if $G$ has a vertex cover of
size at most $k$ then deleting the candidates that correspond to this
vertex cover guarantees $p$'s victory. This completes the proof for the
nonunique-winner case.

To obtain the proof for the unique-winner case, we need to add one
more candidate, $\hat{r}$, that is
a ``clone'' of $r$ (i.e., $\hat{r}$ ties in
the head-to-head contest with $r$ and has the same results as $r$ in all
other head-to-head contests). In such a modified election, $p$
has the same $\copelandalpha$ score as each of the $e_i$'s and has to
gain at least one point over each of them to become the unique
winner. The rest of the argument remains the same.~\end{proofs}

Theorem~\ref{thm:ccdc} will be helpful in treating the (run-off)
par\-tition-of-candidates cases.  Again, it is known
from~\cite{fal-hem-hem-rot:c-ABBREVIATE:llull} that both Copeland$^0$
and Copeland$^1$ are resistant to constructive control by (run-off)
partition of candidates in both the ties-promote model and the
ties-eliminate model.

\begin{theorem}
\label{thm:ccrpc}
Let $\alpha$ be a rational number such that $0 \leq \alpha \leq 1$.
$\copelandalpha$ is resistant to constructive control via run-off
partition of candidates in both the ties-promote model (CCRPC-TP) and the
ties-eliminate model (CCRPC-TE).
\end{theorem}

\sproof
Our proof will, again, follow via a reduction from the vertex cover
problem. Our input is a graph $G$ and a nonnegative integer $k$ and we
seek an election $E$ where our favorite candidate $p$ can be made a
winner in the CCRPC-TP (respectively, CCRPC-TE) model
if and only if $G$ contains a vertex cover of size at most~$k$.
Let $G$ have the edge set $E(G) = \{e_1, \ldots,e_m\}$ and the 
vertex set $V(G) = \{1, \ldots, n\}$.
The following construction is the basis of our proof.

\begin{construction}
  \label{con:ghr}
  Let $F$ and $H$ be two elections,
 with candidate sets $\{f_1, \ldots, f_n\}$ and
  $\{h_1, \ldots, h_q\}$,
respectively.
Define $E = (C,V)$,
  where $C = \{r, f_1, \ldots, f_n, h_1, \ldots, h_q\}$ and voters in
  $V$ are set so that we have the following results of the
  head-to-head contests:
(1) for each $f_i \in C$, $f_i$ defeats $r$;
(2) for each $h_i \in C$, $r$ defeats $h_i$;
(3) for each $h_i, f_j \in C$, $h_i$ defeats $f_j$;
(4) all the remaining head-to-head contests are as in $F$ and
      $H$, respectively.
\end{construction}

Intuitively, Construction~\ref{con:ghr} works as follows.  We set $F$
to be an election that contains the candidate $p$ and where $p$ can be
made a winner (respectively, 
the unique winner) via deleting a set $D$ of at most $k$ candidates
(specifically---and importantly---we will use the elections built in
the proof of Theorem~\ref{thm:ccdc}). We will set the election $H$ in
such a way that the only run-off partitions of candidates in $E$ that
could possibly result in $p$ being a winner (respectively, 
the unique winner) would
form two subcommittees such that the first one would contain
candidates in $F$, possibly without having up to $k$ of them, and the
other subcommittee would contain $r$, the candidates from $H$, and the
remaining candidates in $D$. This way the problem of constructive
control via run-off partition of candidate reduces to the problem of
finding the set $D$, which in Theorem~\ref{thm:ccdc} we have shown to
be $\np$-complete.

Let $F$ and $H$ be two elections where $H$ contains at least
two candidates, and let $E$ be the election obtained from $F$
and $H$ using Construction~\ref{con:ghr}.  We assume that our
preferred candidate, $p$, belongs to $F$, and we assume that there are
no ties in head-to-head contests between candidates in $H$.
Later on we will precisely
specify how the elections $F$ and $H$ are built, and for now we only
mention that in the TE case we will have $H$ have a unique winner.

We have the following result regarding the possible structure
of the subelections in the run-off partition of candidates.

\begin{lemma}
  \label{thm:ccrpc-structure}
  Let $(C_1,C_2)$ be a partition of candidates in $E$ such that $p$ is
  a winner (respectively, 
  the unique winner), where $p$ participates in subelection
  $C_1$.  It holds that $C_1 = F - D$ and $C_2 = H \cup D \cup \{r\}$,
  where $D \subseteq F - \{p\}$.
\end{lemma}
Lemma~\ref{thm:ccrpc-structure} follows directly from
Lemma~\ref{thm:ccrpc-subcom}
below the proof of which is omitted.

\begin{lemma}
  \label{thm:ccrpc-subcom}
  Let $(C_1,C_2)$ be a partition of candidates in $E$ such that $p$
  is a winner (respectively, the unique winner). The subcommittee that contains
  $p$ does not contain any member of $H$ nor~$r$.
\end{lemma}
\OMIT{
\sproof
The proof follows via a case-by-case analysis.

\begin{description}
\item[Case~1:] $p$ is in the same subcommittee as at least two
candidates from $H$, call them $h$ and~$h'$.  Then $p$ is not a winner of
this subelection. The reason is that $p$ wins head-to-head contests
only with candidates from $F$ and possibly with $r$. On the other
hand, both $h$ and $h'$ win head-to-head contests with all members of
$F$ and either $h$ defeats $h'$ or the other way round. As a result,
either $h$ or $h'$ has a $\copelandalpha$ score strictly higher than
that of $p$.

\item[Case~2:] $p$ is in the same subcommittee as exactly one
member of $H$,
say~$h$. It is easy to see, via an analysis
similar to the one above, that if $p$'s subcommittee does not contain
$r$ then $p$ has a $\copelandalpha$ score lower than $h$ and so is not a
winner of the subelection. Thus, let us assume that $r$ participates
in $p$'s subcommittee. In this case, $p$ has
a $\copelandalpha$ score at
most as high as that of $h$.  If $p$ is not a winner or if we are in the
TE model, then we are done. On the other hand, if both $p$ and $h$ (and
possibly $r$) are winners and we are in the TP model then both $p$ and
$h$ will meet the winners of the other subcommittee in the final
round.  However, it is easy to see that the winner of the other
subcommittee has to be some other member of $H$,
say~$h'$. Thus, the final round involves at least $p$, $h$, and $h'$, and via
the arguments from Case~1 we know that $p$ cannot be a winner of such a
final round.
\end{description}

This completes the proof.~\eproofof{Lemma~\ref{thm:ccrpc-subcom}}
} %

\OMIT{
\begin{lemma}
  \label{thm:ccrpc-r}
  Let $(C_1,C_2)$ be a partition of candidates in $E$ such that $p$
  is a winner (respectively, 
  the unique winner). The subcommittee that contains
  $p$ does not contain $r$.
\end{lemma}
\sproof
Without loss of generality,
we assume that $p$ is in subcommittee $C_1$.  Note that via
Lemma~\ref{thm:ccrpc-subcom} we know that $C_1$ does not contain any
members of $H$. Assume, for the sake of contradiction, that $r \in
C_1$. Since $r$ loses to all members of $F$, $r$ is certainly not a
winner of subelection $C_1$. It is also easy to see that winners of
$C_2$ are members of $H$ (the winner of $C_2$ is a member of
$H$). Thus, even if $p$ makes it to the final round, he or she meets
there with some $h$ from $H$ who has a higher $\copelandalpha$ score
than $p$: $h$ defeats $p$ in their head-to-head contest as well as
every other candidate that $p$ defeats in head-to-head contests ($r$ is
not participating in the final round).  Thus, $p$ is not a winner
(respectively, is not the
unique winner); a contradiction. The lemma is
proven.~\eproofof{Lemma~\ref{thm:ccrpc-r}}
} %

Using the above lemmas, we can specify the
elections $F$ and $H$ and complete the proof.  We will first handle the
nonunique-winner cases and then we will argue how to modify the proof
to apply to the unique-winner model.

For the ties-promote (respectively, ties-eliminate) case, we set $F$
to be the election built in the proof of Theorem~\ref{thm:ccdc} for
the nonunique-winner model (respectively, for the unique-winner
model).  For the ties-promote case, we set $H$ to be an election with
candidate set $\{r, h_1, \ldots, h_q\}$ such that, for some
nonnegative integer $\ell$, we have the following
scores:\footnote{Candidate $r$ in Construction~\ref{con:ghr} was,
strictly speaking, not a member of $H$, but for the sake of building
the election $E$ here it is easier to consider members of $H$ and $r$
jointly.}
(a) $\copelandalphascore_H(r) = \ell$;
(b) $\copelandalphascore_H(h_1) = \ell - k - 1$;
(c) $\copelandalphascore_H(h_2) = \ell - k - 1$;
(d) for each $i \in \{3, \ldots, q\}$,
 $\copelandalphascore_H(h_i) \leq \ell - k - 1$.
Such an election is easy to build in polynomial time using
Lemma~\ref{thm:construction-lemma}. 
For the ties-eliminate case, we set $H$ to have candidate set
 $\{r, h_1, \ldots, h_q\}$
with the following scores:
(a) $\copelandalphascore_H(r) = \ell$;
(b) $\copelandalphascore_H(h_1) = \ell - k$;
(c) for each $i \in \{2, \ldots, q\}$,
 $\copelandalphascore_H(h_i) < \ell - k$.

\begin{lemma}
  \label{thm:ccrpc-k-bound}
  Set $D$ in Lemma~\ref{thm:ccrpc-structure} cannot contain more than
  $k$ elements.
\end{lemma}
\OMIT{
\sproof
From Lemma~\ref{thm:ccrpc-structure} we know that we have a
subcommittee $H \cup D \cup \{r\}$. Note that if $\|D\| > k$ then in
the TP case this subcommittee has two winners, $h_1$ and $h_2$. Even if
$p$ would be promoted to the final round, he or she would meet two
members of $H$ there and would not become a global winner. If $\|D\| > k$ in
the TE case then $h_1$ would be the unique winner of this
subcommittee. If $p$
were the unique winner of his or her subcommittee then the final round
would be between $p$ and $h_1$, and $p$ would
lose.~\eproofof{Lemma~\ref{thm:ccrpc-k-bound}}
} %

The proof of Lemma~\ref{thm:ccrpc-k-bound} is omitted.
To complete the proof of Theorem~\ref{thm:ccrpc}, note that
since $p$ can become a winner of his or her subcommittee if and only
if $p$ can be made a winner (respectively, 
the unique winner) of election $F - D$,
where $D \subseteq F - \{p\}$ and $\|D\| \leq k$, we see that $p$ can
become a winner (respectively, 
the unique winner) only if $G$ has a vertex cover of
size at most $k$.  This follows by our choice of~$F$.
On the other hand,
if we choose $D$ to be the
set of vertices that correspond to an at-most-size-$k$ vertex cover of $G$
and partition the candidates
in $C$ as in Lemma~\ref{thm:ccrpc-structure}, then $p$ is a winner
(respectively, the unique winner),
since if we use such a partition and the set
$D$ then the subcommittee $H \cup D \cup \{r\}$ either has no winner or
has the unique winner~$r$, and the subcommittee $F - D$ either has the
unique winner $p$ (in the TE case) or has winner set $\{p\} \cup W$
(in the TP case), where $W \subseteq \{e_1, \ldots, e_m\}$.
Since $p$ and all members
of $W$
are tied, they all become the winners of election $E$. This
completes the proof for the nonunique-winner case.

We still need to handle the unique-winner cases. However, note that in
the case of the TE model the current proof already works just as well
in the unique-winner model. It remains to handle the unique-winner case
in the ties-promote model.
However, in this case we simply
need to take $F$ to be the unique-winner version of election $E$ from
Theorem~\ref{thm:ccdc}.  Our lemmas describing the structure of the
subcommittees apply in the ties-promote case, and so if $p$ is to be a
winner then $r$ should be the unique winner of subcommittee $H \cup
D \cup \{r\}$, where $D$ is a subset of $F - \{p\}$ with at most $k$
elements, and the global winner of the election, if any, is the unique
winner of election $F - D$. Since we know that $p$ can become a unique
winner of $F$ via deleting at most $k$ candidates if and only if $G$
has a vertex cover of size at most $k$, the proof is
complete.~\eproofof{Theorem~\ref{thm:ccrpc}}

Finally, we state without proof our result for constructive control by
partition of candidates, which extends the corresponding known result
for Copeland$^0$ from~\cite{fal-hem-hem-rot:c-ABBREVIATE:llull}.
Unlike in the case of run-off partition of candidates, however, our
proof does not apply to the case of $\alpha = 1$ for the
TE model, but note that these resistances of Copeland$^1$
were already shown in~\cite{fal-hem-hem-rot:c-ABBREVIATE:llull} for
both the
TP and the
TE model.

\begin{theorem}
\label{thm:ccpc-tp-te}
\begin{enumerate}
\item Let $\alpha$ be a rational number with $0 \leq \alpha \leq 1$.
  $\copelandalpha$ is resistant to constructive control via partition
  of candidates in the ties-promote model (CCPC-TP).
\item Let $\alpha$ be a rational number with $0 \leq \alpha < 1$.
  $\copelandalpha$ is resistant to constructive control via partition
  of candidates in the ties-eliminate model (CCPC-TE).
\end{enumerate}
\end{theorem}

\subsection{Voter Control}
\label{sec:control-voter}

Our first result regarding voter control extends to all rationals
$\alpha$, $0 \leq \alpha \leq 1$, the corresponding result for
Copeland$^0$ and Copeland$^1$
from~\cite{fal-hem-hem-rot:c-ABBREVIATE:llull}.  The proof is omitted.

\begin{theorem}
\label{thm:ccav-dcav}
  Let $\alpha$ be a rational number such that $0 \leq \alpha \leq
  1$. $\copelandalpha$ is resistant to both constructive and
  destructive control via adding voters (CCAV and DCAV).
\end{theorem}

Next, we state without proof our result for control by deleting
voters, which extends the corresponding known result for Copeland$^0$
from~\cite{fal-hem-hem-rot:c-ABBREVIATE:llull}.
\jfootnote{Edith writes
  in email of Sun, Oct 7, 08:27:57 pm: ``Looks to me like the
  construction (and the proof) of Theorem 4.27 (CCDV, $\alpha < 1$) is
  wrong.  After deleting the voters corresponding to the cover, $p$
  *ties* every other candidate.  And, in fact, this is the
  construction for *Llull* from my notes (there is a slightly modified
  construction in my notes for non-Llull).''}
Note that our proof does not apply to the case of $\alpha = 1$, but we
mention that these resistances of Copeland$^1$ were already shown
in~\cite{fal-hem-hem-rot:c-ABBREVIATE:llull}.

\begin{theorem}
\label{thm:ccdv-dcdv-smaller-than-one}
  Let $\alpha$ be a rational number such that $0 \leq \alpha < 1$.
  $\copelandalpha$ is resistant to both constructive and destructive
  control via deleting voters (CCDV and DCDV).
\end{theorem}

Finally, we state without proof our result for control
by (run-off) partition of voters,
\jfootnote{CHECK:
  Theorem~\ref{thm:control-voter--constructive-destructive} and its
  proof!}
which extends the corresponding results for Copeland$^0$ and
Copeland$^1$ from~\cite{fal-hem-hem-rot:c-ABBREVIATE:llull}.

\begin{theorem}
\label{thm:control-voter--constructive-destructive}
  Let $\alpha$ be a rational number such that $0 \leq \alpha \leq 1$.
  $\copelandalpha$ is resistant to constructive and destructive control
  via partition of voters (in both the TP and TE model, i.e.,
  CCPV-TP, CCPV-TE, DCPV-TP, and DCPV-TE), and to run-off partition
  of voters (in both the TP and TE model, i.e., CCRPV-TP, CCRPV-TE,
  DCRPV-TP, and DCRPV-TE).
\end{theorem}

\subsection{FPT Algorithm Schemes for Bounded-Case Control}
\label{sec:fpt-bounded-case-control}

\subsubsection{Fixed-Parameter Tractability Results}
\label{sec:fpt}

In their seminal paper on NP-hard winner-determination problems,
Bartholdi, Tovey, and Trick~\cite{bar-tov-tri:j:who-won} suggested
considering hard election problems for the cases of a bounded number
of candidates or a bounded number of voters, and they obtained
efficient-algorithm results for such cases.  Within the study of
elections, this same approach---seeking efficient
fixed-parameter algorithm families---has
also been used, for example,  within the study of
bribery~\cite{fal-hem-hem:c-ABBREVIATE:bribery}.
Faliszewski et al.~\cite{fal-hem-hem-rot:c-ABBREVIATE:llull} showed
that the 16 resistance results for constructive and destructive voter
control within Copeland$^0$ and Copeland$^1$ (see
Table~\ref{tab:control}) are in FPT (i.e., they each are
fixed-parameter tractable) if the number of candidates is bounded,
and also if the number of voters is bounded.  They also showed
that these results hold even when the multiplicities of
preference lists in a given election are represented succinctly (by a
binary number).  

We extend these results in Theorems~\ref{t:v} and~\ref{t:c-bc} below.
To state these results concisely, we
borrow a notational approach from transformational grammar, and use
square brackets as an ``independent choice'' notation.  So, for
example, the claim \mbox{$\scriptsize
\begin{bmatrix} \textrm{\rm{}It} \\ \textrm{\rm{}She} \\ \textrm{\rm{}He} \end{bmatrix}
\begin{bmatrix} \textrm{\rm{}runs} \\ \textrm{\rm{}walks} \end{bmatrix}
$} %
is a shorthand for six assertions: It runs; She runs; He runs;
It
walks; She walks; and He 
walks.  
A special case is the symbol
``$\emptyset$'' which, when it appears in such a bracket, means that
when unwound it should be viewed as no text at all.  For example,
``$\bigl[\begin{smallmatrix} \textrm{\rm{}Succinct} \\ \emptyset
\end{smallmatrix}\bigr]$ Copeland is fun'' asserts both ``Succinct 
Copeland is fun'' and ``Copeland is fun.''

\begin{theorem}\label{t:v}
For each rational $\alpha$, $0\leq \alpha \leq 1$, and each 
choice from the independent choice brackets below, 
the specified problem family
(as $j$ varies over $\naturals$) is in $\fpt$:
$\begin{bmatrix}
\textrm{\rm{}succinct}\\\textrm{\rm{}$\emptyset$}
\end{bmatrix}
\hbox{-}
\begin{bmatrix}
\textrm{\rm{}\copelandalpha}\\\textrm{\rm{}\copelandalphairrational}
\end{bmatrix}
\hbox{-}
\begin{bmatrix}
\textrm{\rm{}C}\\\textrm{\rm{}D}
\end{bmatrix}
\hbox{\rm{}C}
\begin{bmatrix}
\textrm{\rm{}AV}\\
\textrm{\rm{}DV}\\
\textrm{\rm{}PV-TE}\\
\textrm{\rm{}PV-TP}
\end{bmatrix}
\hbox{-}
\begin{bmatrix}
\textrm{\rm{}\bvj}\\\textrm{\rm{}\bcj}
\end{bmatrix}$.
\end{theorem}

\begin{theorem}\label{t:c-bc}
For each rational $\alpha$, $0\leq \alpha \leq 1$, and each 
choice from the independent choice brackets below, the specified
problem family
(as $j$ varies over $\naturals$) is in $\fpt$:
$\begin{bmatrix}
\textrm{\rm{}succinct}\\\textrm{\rm{}$\emptyset$}
\end{bmatrix}
\hbox{-}
\begin{bmatrix}
\textrm{\rm{}\copelandalpha}\\\textrm{\rm{}\copelandalphairrational}
\end{bmatrix}
\hbox{-}
\begin{bmatrix}
\textrm{\rm{}C}\\\textrm{\rm{}D}
\end{bmatrix}
\hbox{\rm{}C}
\begin{bmatrix}
\textrm{\rm{}AC${}_{\rm u}$}\\
\textrm{\rm{}AC}\\
\textrm{\rm{}DC}\\
\textrm{\rm{}PC-TE}\\
\textrm{\rm{}PC-TP}\\
\textrm{\rm{}RPC-TE}\\
\textrm{\rm{}RPC-TP}
\end{bmatrix}
\hbox{-$\bcj$}$.
\end{theorem}

The proofs of Theorems~\ref{t:v} and~\ref{t:c-bc}, which in particular
employ Lenstra's~\cite{len:j:integer-fixed}
algorithm for bounded-variable-cardinality integer programming, are
omitted here.

\subsubsection{FPT and Extended Control}

In this section, we introduce 
and look at extended control.  By that we do not mean
changing the
basic control notions of
adding/dele\-ting/partitioning candidates/voters.  Rather, we mean
generalizing past merely looking at the constructive (make a
distinguished candidate a winner) and the destructive (prevent a
distinguished candidate from being a winner) cases.  In particular, we
are interested in control where the goal can be far
more flexibly specified, for example (though in the partition cases
we will be even more flexible than this), we will allow 
as our goal region any
(reasonable---there are some time-related conditions) subcollection of
``Copeland outcome tables'' (specifications of who won/lost/tied each
head-to-head contest).

Since from a Copeland outcome
table, in concert with
the current $\alpha$, one can read off the $\copelandalphairrational$ scores of
the candidates, this allows us a tremendous range of descriptive
flexibility in specifying our control goals, e.g., we can specify a
linear order desired for the candidates with respect to their
$\copelandalphairrational$ scores, we can specify a linear-order-with-ties
desired for the candidates with respect to their $\copelandalphairrational$
scores, we can specify the exact desired $\copelandalphairrational$ scores for
one or more candidates, we can specify that we want to ensure that
no candidate from a certain subgroup has a $\copelandalphairrational$ score that
ties or beats the $\copelandalphairrational$ score of any candidate from a
certain other subgroup, etc.

All the FPT algorithms given in the previous section regard, on their
surface, the standard control problem, which tests whether a given
candidate can be made a winner (constructive case) or can be precluded
from being a winner (destructive case).  We
note that the
general approaches used in that section in fact yield 
FPT schemes even for the far more
flexible notions of 
control mentioned above.

\subsubsection{Resistance Results}\label{ss-star:resistance}

In contrast with the FPT results
in~\cite{fal-hem-hem-rot:c-ABBREVIATE:llull} for Copeland$^0$ and
Copeland$^1$, Faliszewski et
al.~\cite{fal-hem-hem-rot:c-ABBREVIATE:llull} showed that, for $\alpha
\in \{0,1\}$, $\copelandalphairrational$ remains resistant to all
types of candidate control even for two voters.  We extend these
results by showing that even for each rational~$\alpha$, $0\leq \alpha
\leq 1$, for $\copelandalphairrational$ all 19 candidate-control
cases that we showed earlier in this paper (i.e., without bounds on
the number of voters) to be resistant remain resistant even for the
case of bounded voters (nonsuccinct).  This resistance holds even when the input is
not in succinct format, and so it certainly also holds when the input
is in succinct format.

It remains open whether Table 1's resistant, rational-voter,
candidate-control cases remain resistant for the bounded-voter case.

\section{Bribery}
\label{sec:bribery}

Theorem~\ref{thm:bribery} extends to all rationals $\alpha$, $0 \leq \alpha
\leq 1$, the corresponding result for Copeland$^0$ and Copeland$^1$
from~\cite{fal-hem-hem-rot:c-ABBREVIATE:llull}.

\begin{theorem}
\label{thm:bribery}
For each rational~$\alpha$, $0 \leq \alpha \leq 1$,
$\copelandalpha$ is resistant to both constructive and
  destructive bribery in
  both the rational-voters case and the irrational-voters case.
\end{theorem}

We also extend another result for Copeland$^0$ and Copeland$^1$
from~\cite{fal-hem-hem-rot:c-ABBREVIATE:llull} to all rationals
$\alpha$, $0 \leq \alpha \leq 1$: $\copelandalphairrational$ is
vulnerable to destructive microbribery.  Informally put, microbribery
means that the briber pays separately for each preference-table entry
flip of irrational voters.

All proofs of Sections~\ref{sec:fpt-bounded-case-control}
and~\ref{sec:bribery} are omitted due to space, but can be found
(along with other omitted proofs) in the currently 82-page,
in-preparation full version of this paper.

\section{Conclusions}

In this paper we studied $\copelandalpha$ elections with respect to
their resistance and vulnerability to control and bribery.  Among the
election systems whose winners can be determined in polynomial time,
we identified the first natural election system, Copeland 
(i.e., Copeland$^{0.5}$),
that provides full resistance to constructive control.  Using this
result, we also obtained
the first (hybrid) election system that is resistant to each type of
constructive and destructive control, has a polynomial-time winner
problem, and is built only from natural election systems.  In
addition, we extended previous resistance results
on bribery and fixed-parameter tractability of bounded-case
control to $\copelandalpha$ for each rational $\alpha$, $0 < \alpha <
1$.  Regarding the latter, questions that remain open concern
the rational-voter, candidate-control, bounded-voter cases.  Another
open question regards the complexity of constructive microbribery for
Copeland$^{0.5}_{\rm{}Irrational}$.

{\small 

\bibliographystyle{abbrv}

\bibliography{gryaamas}  

}
\end{document}